\documentclass[footinbib,aps,pra,reprint,superscriptaddress,longbibliography]{revtex4-1}
\usepackage{graphicx}
\usepackage{amsmath}
\usepackage[alsoload=synchem]{siunitx}
\usepackage{braket}
\usepackage{epstopdf}
\usepackage{url}

\newcommand{\reffig}[1]{figure \ref{#1}}
\newcommand{\refsubfig}[2]{figure \ref{#1}(#2)}

\begin{document}

\title{High-efficiency cold-atom transport into a waveguide trap}

\author{A. P. Hilton}
\email[]{ashby.hilton@adelaide.edu.au}
\author{C. Perrella}
\affiliation{Institute for Photonics and Advanced Sensing (IPAS) \& School of Physical Sciences, The University of Adelaide, Adelaide, South Australia 5005, Australia}
\author{F. Benabid}
\affiliation{GPPMM group, Xlim Research Institute, UMR CNRS 7252, Universit\'e de Limoges, France}
\author{B. M. Sparkes}
\author{A. N. Luiten}
\author{P. S. Light}
\affiliation{Institute for Photonics and Advanced Sensing (IPAS) \& School of Physical Sciences, The University of Adelaide, Adelaide, South Australia 5005, Australia}

\date{\today}

\begin{abstract}
	We have developed and characterized an atom-guiding technique that loads \num{3e6} cold rubidium atoms into hollow-core optical fibre, an order-of-magnitude larger than previously reported results.
	This result was possible because it was guided by a physically realistic simulation that could provide the specifications for loading efficiencies of \SI{3.0}{\percent} and a peak optical depth of \num{600}.
	The simulation further showed that the demonstrated loading efficiency is limited solely by the geometric overlap of the atom cloud and the optical guide beam, and is thus open to further improvement with experimental modification. 
	The experimental arrangement allows observation of the real-time effects of light-assisted cold atom collisions and background gas collisions by tracking the dynamics of the cold atom cloud as it falls into the fibre.
	The combination of these observations, and physical understanding from the simulation, allows estimation of the limits to loading cold atoms into hollow-core fibres.
\end{abstract}

\pacs{}

\maketitle

\section{Introduction}
	Cold atoms have been a transformational tool for sensing \cite{Kasevich1991,Dutta2016,Canuel2006}, measurement \cite{Peters2001,Stockton2011,Hinkley2013,Cronin2009,Altin2013}, emulation \cite{Glaetzle2017} and simulation \cite{Gross2017,Garreau2017}.
	The specific properties of cold matter that makes it useful for these applications is its high atomic density, low velocity, and excellent isolation from the environment.
	Furthermore, alkali metals are commonly used in cold matter experiments as they confer strong atom-light interaction allowing both efficient measurement and manipulation.
	The  figures of merit that quantify these properties for  an atomic ensemble are its coherence time, $\tau$, and optical depth, defined as $\mathcal{D}^\mathrm{opt} = -\ln{(T)}$, where $T$ is the transmission.
	Examples where overall performance depends crucially on these figures of merit are quantum state storage \cite{Kasevich1991,Sparkes2013}, strong photon-photon interaction for quantum information processing \cite{Liu2016}, and interferometric magnetic gradiometry \cite{Hardman2016}.
	
	A promising approach to attain a high $\mathcal{D}^\mathrm{opt}$ is the loading of cold atoms into a hollow-core photonic crystal fibre (HC-PCF) \cite{Cregan1999,Couny2006,Markos2017}. 
	The tight confinement of atoms to the core of the fibre delivers a close match between the optical cross-section of the cold atom and the transverse mode diameter of the light field.
	The close matching leads to an optimal $\mathcal{D}^\mathrm{opt}$ for a given atom number.
	Further, the guidance of the fibre means the match can be extended over lengths that are not limited by diffraction - up to \num{10}s of centimetres.
	
	Previous experiments aimed at loading cold atoms into HC-PCF have achieved $\mathcal{D}^\mathrm{opt}$ up to \num{1000} \cite{Blatt2014} and have been used to demonstrate slow and stopped light \cite{Blatt2016} as well as highly efficient few-photon all-optical switching \cite{Peyronel2012} and excitation of Rydberg atoms \cite{Langbecker2017}.
	Similar systems have shown coherence times not limited by atom-wall or atom-guide interactions using Lamb-Dicke spectroscopy in a 1-D lattice \cite{Okaba2014}, or atom interferometry using optically-confined free-falling cold atoms \cite{Xin2018}.
	
	These preliminary explorations show promising results; however, the limitations and dynamics of the atomic loading and trapping process are still not well understood.
	This work provides a detailed simulation of the cold atom loading of an optical fibre, which is augmented with an innovative experimental design that can  follow the   atoms during the loading and trapping process.
	The simulation is seen to be in excellent agreement with the experiment and the imbued confidence allows rapid pinpointing of the optimal conditions in which more than \num{3} million cold atoms can be loaded into the fibre.
	Using the validated simulation we are able to predict the limits to optical depth and coherence lifetimes and suggest techniques to expand the capacity of this platform for implementing coherent state storage and manipulation.
	
\section{Methods}
	\label{sec:examples}
	
\begin{figure}
		\centering
		\includegraphics[]{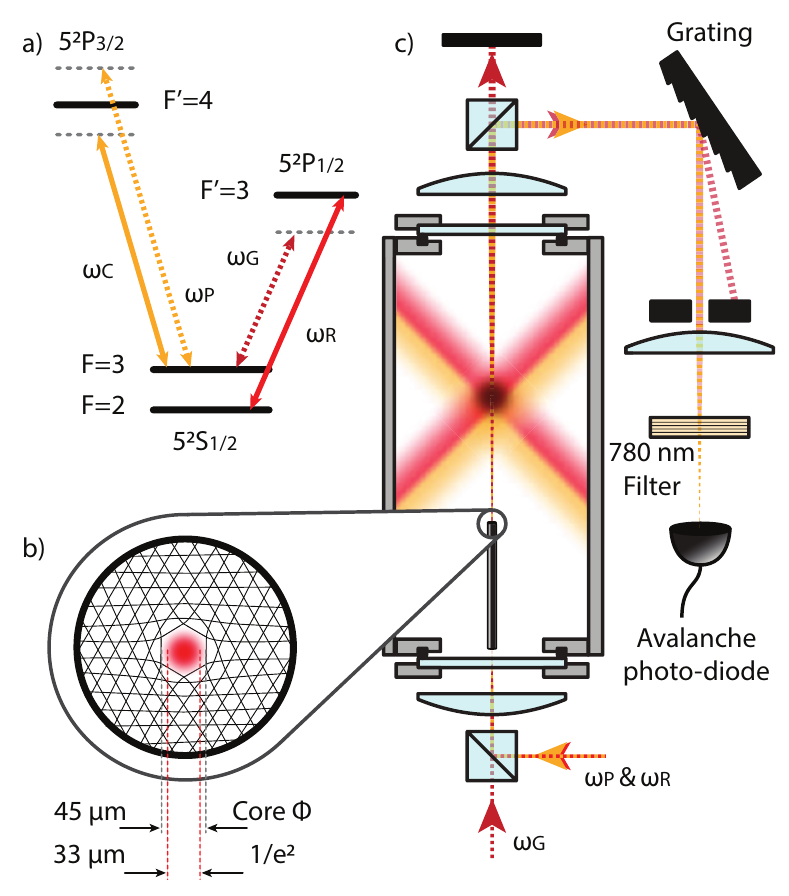}
		\caption{
			Experimental setup with 
			a) the relevant energy levels in the D1 \& D2 transitions, along with the relative frequency of the MOT cooling laser $\omega_\mathrm{C}$, the repump laser $\omega_\mathrm{R}$, the guide laser $\omega_\mathrm{G}$, and the probe laser $\omega_\mathrm{P}$,
			b) a cross section of the Kagome HC-PCF, and 
			c) a schematic of the fibre coupling system.
			\label{fig:SetupFig}
		}
	\end{figure}

	The experiment uses cold rubidium-85 atoms produced in a standard 3-dimensional magneto optical trap (MOT).
	The atoms are cooled on the $F=3 \rightarrow F'=4$ cycling transition of the D$_2$ line, and repumped on the $F=2 \rightarrow F'=3$ transition of the D$_1$ line, shown in \refsubfig{fig:SetupFig}{a}.
	We trap \num{1e8} atoms in a cloud \SI{2}{\milli\meter} in diameter with a steady state temperature of \SI{150}{\micro\kelvin}.
	The ensemble temperature is reduced below \SI{5}{\micro\kelvin} using $\sigma_+\sigma_-$ polarisation gradient cooling (PGC) \cite{Dalibard1989}.
	A \SI{10}{\centi\meter} length of \SI{45}{\micro\meter} core diameter Kagome-lattice fibre (see \refsubfig{fig:SetupFig}{b}) is situated \SI{25}{\milli\meter} under the MOT as shown in \refsubfig{fig:SetupFig}{c}.  
	This fibre has low loss between \SI{600}{\nano\meter} and \SI{1600}{\nano\meter}, and we are able to achieve a combined guidance and coupling efficiency into the fundamental mode of over \SI{70}{\percent} at \SI{780}{\nano\meter}.
	After the PGC phase an intense guide laser \cite{Grimm2000} is switched on producing trapping forces that cause the falling atoms within the laser field to be guided towards the core of the fibre.
	The guide beam is coupled into the fibre from below, which produces an attractive dipole trap within the hollow-core fibre that diverges from the output to produce a conical, self-aligning optical funnel that steers atoms into the fibre.
	With over \SI{1}{\watt} of guide light detuned by \SI{1}{\tera\hertz} to the red of the D$_1$ transition we produce a radial trapping field with a peak trap depth $U_{\mathrm{dip}}$ of \SI{20}{\micro\kelvin} at the MOT location without the need for intricate magnetic field guides \cite{Bajcsy2011}.
	The eventual fraction of atoms guided into the fibre  depends on the velocity and density distributions of the initial ensemble as well as its position with respect to the guide beam.
	These are complicated functions of the MOT and PGC parameters as well as the settings of the background magnetic field cancellation coils.
	We maximize the loading dependence on the initial state of the MOT by placing \num{10} of the most sensitive of these variables under the control of a neural-net learning tool, M-Loop \cite{Wigley2016,HushMLoop}, which is capable of efficiently optimising over this multidimensional space.

	The atom number contained within the guide beam is estimated using the transmission of a weak (\SI{5}{\nano\watt}) co-propagating light field that is tuned around the D$_2$ transition.
	The transmitted probe is separated from the guide using high extinction polarisation optics, an optical grating, and finally a narrow band-pass filter, which suppresses the guide light by \SI{90}{\decibel}.
	Prior work \cite{Bajcsy2009,Blatt2014,Okaba2014,Xin2018} has implemented a spatial-mode filter of the probe light on exit from the fibre: we did not  do that here as strong lensing effects within the large fibre core \cite{Noaman2018,Roof2015} would lead to an over-estimate of the optical depth.	
	
	The intense guide light introduced an inhomogeneous broadening of the probe transition through  its transverse intensity variation \cite{Davidson1995}.
	This effect was circumvented by probing in the dark: the guide beam was rapidly intensity modulated and probing was only performed in the absence of the guide light.
	We used two double-passed acousto-optic modulators (AOMs) in the probe path to perform the fast (\SI{30}{\nano\second}) switching required.
	This fast switching speed delivered us an additional advantage in that we could step the optical frequency between each probing phase, thereby observing the full spectral width of the absorption feature over \SI{144}{\mega\hertz} in a cycle lasting just \SI{100}{\micro\second}. A small amount of re-pump light is present with probe, ensuring that population is not lost to the $F=2$ ground state during measurement.
	We confirmed that our technique was non-destructive by applying two consecutive probe sequences and measuring less than \SI{10}{\percent} variation between measurements.
	During each cycle of the experiment we probe the cold cloud with one laser pulse sequence at a pre-determined time following its release from the MOT;  a second identical pulse sequence is then applied \SI{500}{\milli\second} following  the first.
	This second pulse sequence is used to normalize the cold atom absorption as it is  sufficiently delayed that it sees only a small residual absorption associated with the background of hot rubidium atoms - less than \SI{1}{\percent}.
	
\section{Atom Loading Results}
	\begin{figure}
		\centering
		\includegraphics{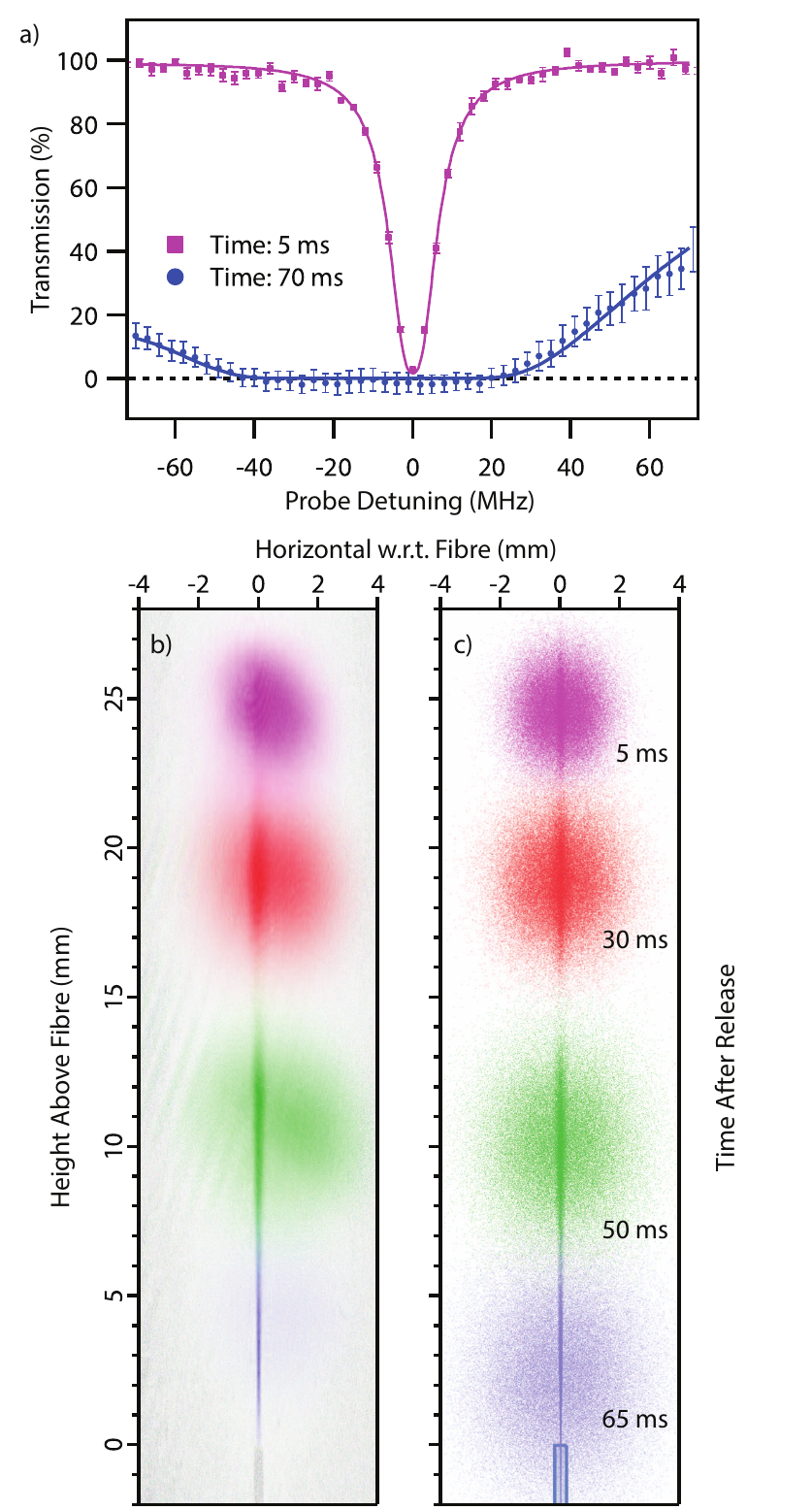}
		\caption{
			Atom loading with \SI{1}{\watt} of guide power detuned by \SI{1}{\tera\hertz} below the $\mathrm{D}_1$ transition, with 
			a) transmission measurements through the fibre at \SI{5}{\milli\second} (purple squares) and \SI{70}{\milli\second} (blue circles) with fits to data,  
			b) optical depth map calculated from absorption imaging from the side at \SI{5}{\milli\second} (purple), \SI{30}{\milli\second} (red), \SI{50}{\milli\second} (green), and \SI{65}{\milli\second} (blue) after release, and
			c) simulated optical depth map that matches experimental conditions at the same instants.
			\label{fig:Filmstrip}
		}
	\end{figure}
	We compare the probe absorption when the atoms are just released from the MOT  (time: \SI{5}{\milli\second}), to that observed as they just enter the fibre (time: \SI{70}{\milli\second}) in \refsubfig{fig:Filmstrip}{a}.
	We see a peak optical depth, $\mathcal{D}^\mathrm{opt}$ of \num{600\pm10}, comparable to the highest depths measured in other fibre-loading experiments \cite{Blatt2014,Blatt2016,Peyronel2012}, and approaching the highest depths possible with other techniques \cite{Sparkes2013,Hsiao2018,Kaczmarek2015,Hsiao2014}.
	
	By calculating the atom number from $\mathcal{D}^\mathrm{opt}$ (see Appendix A) we estimate we have guided \num{3.3 \pm 0.1 e6} atoms into the fibre, which is approximately \SI{3}{\percent} of the MOT cloud.
	Using near-resonance absorption imaging  from the side, see \refsubfig{fig:Filmstrip}{b}, we can directly see the intensifying  atomic density in the guide field as the atoms drop. 
	
	Due to the relatively small diffraction angle of the guide from the fibre only \SI{3.2}{\percent} of the atoms in the initial ensemble  experience a trap depth  larger than the average temperature of the ensemble.
	The Gaussian form of the trap depth in the radial direction from the guide axis, and linear scaling of trap depth with guide power, impose harsh diminishing returns on atom coupling performance for fixed cloud and trap geometries (see Appendix B).
	This strongly suggests that the transfer efficiency could be significantly increased by using density-enhancing techniques in the MOT region e.g. spatial dark spot \cite{Kaczmarek2015,Ketterle1993}, magnetic compression \cite{Depue2000,Petrich1994} or a vertically aligned cigar shaped trap \cite{Lin2008}, or by increasing the divergence of the guide by using a smaller core fibre.

\section{Monte-Carlo Simulation}
	The underlying physics of the atom-light interaction can be modelled using   a detailed Monte-Carlo simulation of the  loading process.
	The simulation picks atoms from within a three dimensional position and velocity distribution  that is matched to the experimentally measured atom-cloud size and temperature.
	The evolution of the atomic position, $\mathbf{r}$, and velocity is then modelled using the differential equation:
	\begin{equation}
		\frac{d^2 \mathbf{r}}{d t^2}=-\frac{\nabla U_\mathrm{dip}(\mathbf{r})}{\mathrm{m}_{\mathrm{Rb}}}+\mathbf{g}
		\label{eqn:EqnMotion}
	\end{equation}
	where $\mathrm{m}_{\mathrm{Rb}}$ is the atomic mass of rubidium-85 and
	\begin{equation}
		U_\mathrm{dip}(\mathbf{r})=\frac{\pi c^2}{2}\left( \frac{\Gamma_{\mathrm{D}_1}}{{\omega_{\mathrm{D}_1}}^3}\frac{1}{\Delta_{\mathrm{D}_1}} + \frac{\Gamma_{\mathrm{D}_2}}{{\omega_{\mathrm{D}_2}}^3}\frac{2}{\Delta_{\mathrm{D}_2}} \right) I(\mathbf{r})
		\label{eqn:TrapDepth}
	\end{equation} 
	describes the potential experienced by a neutral atom in a linearly polarized dipole trap where detunings from the D$_1$ and D$_2$ transitions, $\Delta_{\mathrm{D}_1}$ and  $\Delta_{\mathrm{D}_2}$, are much larger than the ground state hyperfine splitting \cite{Grimm2000} and which the atom is subject to gravitational acceleration, $\mathbf{g}=\SI{-9.81}{\meter\per\second\squared}\hat{z}$.
	Here $\Gamma$ and $\omega$ are the decay rates and optical angular frequencies for specified transitions, and $I(\mathbf{r})$ is the local intensity of the guide beam.
	The simulation includes the spontaneous absorption and emission of guide photons as well as atom loss due to background gas collisions (see Appendix C).
	
	We are able to qualitatively test our simulation by comparing the time dynamics of the ensemble to the experiment.
	A large number ($10^5$) of atoms are simulated, and by calculating the position and optical depth for each atom at several time steps we are able to replicate the data obtained experimentally using both absorption imaging and in-fibre spectroscopy techniques.
	As seen in \refsubfig{fig:Filmstrip}{c} there is excellent agreement between the experiment and simulation.
	It is clear from these images that a large fraction of the initial atom cloud is not guided by the dipole trap, which highlights the importance of improving the mode overlap of the guide and MOT.
	
\begin{figure}
		\centering
		\includegraphics{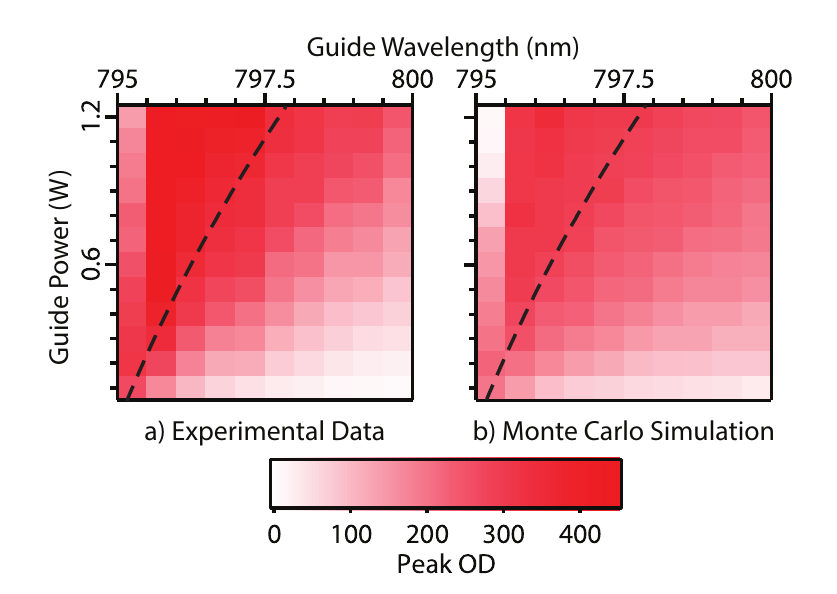}
		\caption{
			Peak optical depth as a fraction of number of atoms in initial MOT 
			a) measured experimentally, and 
			b) calculated using Monte Carlo simulation, 
			both with the line of $U_\mathrm{dip}=\SI{-3}{\micro\kelvin}$ at the cloud (black, dashed).
			\label{fig:EfficiencyPlot}
		}
	\end{figure}

	We experimentally investigated the dependence of the atom loading process on dipole trap parameters by measuring peak optical depth as a function of guide powers and wavelengths.
	These measurements, shown in \refsubfig{fig:EfficiencyPlot}{a}, provide a quantitative means to compare the experiment against the simulation results shown on \refsubfig{fig:EfficiencyPlot}{b}.
	In order to ensure that the experimental atom density distribution is a match for the assumed initial distribution in the simulation we reduced the total atom number density by about a factor of 2 from the values shown before (since it is known that high density MOTs have complex atomic distributions \cite{Walker1990,Gattobigio2010}).
	We see excellent agreement between experiment and simulation, both in the absolute values and the overall shape.
	For conditions in which the guide is nearly resonant (the left-most column in \refsubfig{fig:EfficiencyPlot}{a}) we see a reduction in the number of loaded atoms as we increase the guide depth.
	The simulation provides the physical understanding that this comes because of the high photon scattering rates that drive the atoms away from the fibre.  
	We also see a plateau in the loaded atom number as guide power is increased (marked out by dashed line on the figure) corresponding to a trap deep enough to catch all the atoms in the volume defined by the  overlap of the guide and initial atom ensemble.
	We note that there are no free parameters in the simulation, which emphasizes the strength of the model.

\section{Time Dynamics}
	We have investigated the time dynamics of the loading process by taking absorption measurements at \SI{4}{\milli\second} time intervals following the release of the MOT (similar to that shown in \refsubfig{fig:Filmstrip}{a}.
	Fitting to each spectra provides $\mathcal{D}^\mathrm{opt}$, and we convert this to the equivalent atom number in the fibre.
	To estimate the uncertainty of each measurement and also improve the statistics we have taken \num{5} averages for each time step.
	\begin{figure}
		\centering
		\includegraphics[]{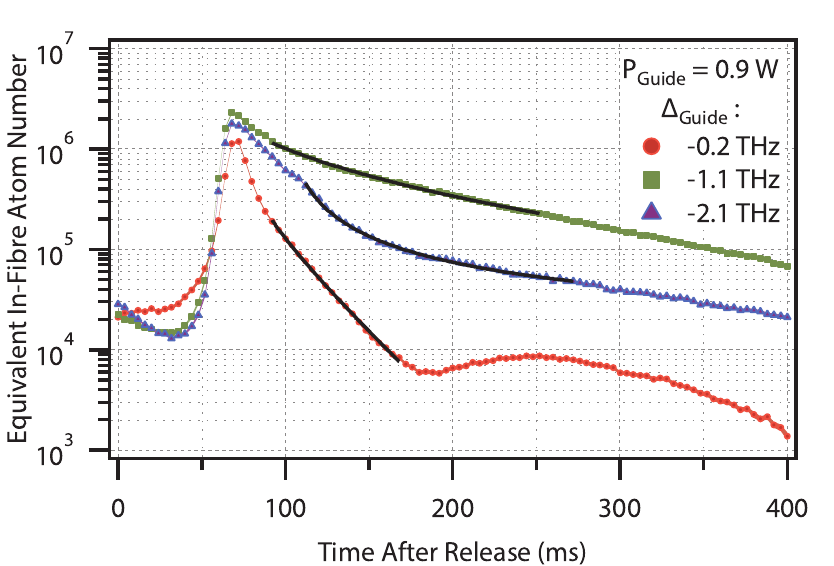}
		\caption{
			Atom number in fibre calculated from optical depth at three different guide detunings.
			The trap depths at the cloud are \SI{150}{\micro\kelvin}, \SI{22}{\micro\kelvin}, and \SI{13}{\micro\kelvin} for detunings of \SI{0.2}{\tera\hertz}, \SI{1.1}{\tera\hertz}, and \SI{2.1}{\tera\hertz} respectively.
			\label{fig:AtomDecay}
		}
	\end{figure}
	
	This procedure is performed in guide detuning steps of $\sim$\SI{200}{\giga\hertz} out to \SI{2.5}{\tera\hertz} below the D$_1$ transition.
	We show in \reffig{fig:AtomDecay} the temporal evolution of atom number for three of these different wavelengths.
	We note that these curves display considerably different loading behaviours.
	For a near-detuned guide we see a very clear peak in the number of atoms in the fibre followed by a rapid decay, and a weak resurgence at \SI{250}{\milli\second}.
	For a \SI{1.1}{\tera\hertz} guide detuning we see a similar dramatic increase in atom number at \SI{\approx 70}{\milli\second} but with a much slower exponential decay of atomic number.  
	For a \SI{2.1}{\tera\hertz} guide detuning, where one might have expected weaker interaction with the guide and thus similar behaviour to the \SI{1.1}{\tera\hertz} result, we  observe an initially higher decay rate.
	At longer times we see a return to an exponential decay with a similar coefficient to the \SI{1.1}{\tera\hertz} detuning situation. 
	Between these specific examples the loading dynamics follow a smooth transition between each regime.
	
	Our simulation allows us to explain these behaviours: for near-detuning the photon scattering rate from the guide is high and atoms are quickly expelled from the fibre leading to the strong decay.
	As seen on \reffig{fig:AtomDecay}, some of the atoms that have been pushed upwards eventually re-enter the fibre at a later time leading to the observation of the resurgence of absorption associated with ``bouncing'' atoms.
	The theoretical treatment allows us to observe these ``bouncing'' atoms in the simulated absorption images \footnote{See Supplementary Material at [URL HERE] for simulation atom bounce absorption images.}.
	For the case with \SI{1.1}{\tera\hertz} detuning, the scattering rate is small enough that we don't see this repulsive effect of the guide.
	In this case, the residual exponential loss is associated with background gas collisions; the observed decay rate allows calculation of the residual background gas density within the fibre (see below).
	For the \SI{2.1}{\tera\hertz} guide detuning we believe that we are driving a photo-association transition resulting in the formation of excited $\mathrm{Rb}_2$ molecules that are lost from the trap \cite{Miller1993}.
	The photo-association rate is proportional to the density of Rb atoms and hence becomes negligible once the density has fallen sufficiently.	
	
	We fit  the atom number evolution for the family of guide detunings to an equation of the form:
	\begin{equation}
		\frac{d N(t)}{dt}=-\gamma N(t)-\beta'(\lambda) N(t)^2
	\end{equation}
	where $\gamma$ is the loss rate due to atomic collisions, and $\beta'(\lambda)$ is the wavelength-dependent photo-association loss coefficient, to extract a collisional loss rate $\gamma$ of \SI{7.6\pm0.2}{\per\second}.
	The derived density-independent loss rate has a similar value to measurements made by \cite{Okaba2014} (\SI{2.9}{\per\second}) in a shorter section of fibre (\SI{32}{\milli\meter}).
	The background rubidium densities in the chamber and the fibre are below $\SI{1e-10}{\torr}$, which results in a Rb-Rb collision rate that is \num{100}-fold too low to explain the observed collision-related loss \cite{Rapol2001}.
	We thus attribute these losses to collisions with background gas atoms - if, as expected, this background is dominated by N$_2$ due to outgassing, then we can calculate the background pressure from the in-fibre collision rate as \SI{1.4\pm0.2e-7}{\torr} \cite{Arpornthip2012,Rapol2001}.
	We can separately calculate the N$_2$ pressure in the chamber from the measured MOT loading time constant giving a value of \SI{2.0\pm0.2e-8}{\torr}.
	The N$_2$ pressure inside the fibre appears reasonable in light of the unfavourable vacuum geometry  of the core of the fibre.
	
\begin{figure}
		\centering
		\includegraphics[]{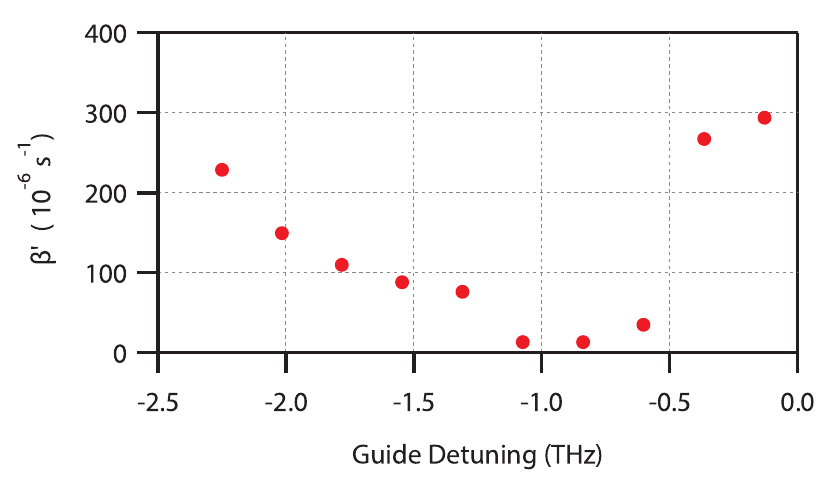}
		\caption{
			Estimation of $\beta'(\lambda)$ from fits to atom decay curves as in \reffig{fig:AtomDecay}.
			\label{fig:bcoef}
		}
	\end{figure}

	The value for $\beta'(\lambda)$ is more difficult to obtain precisely as these processes are dominant during the peak loading time while there are other strongly competing processes.
	We choose a fit window that contains the corner between the fast and slow loss processes (shown as black solid lines in \reffig{fig:AtomDecay}).
	The fit quality is sensitive to the chosen start and end of this fitting window due to the simplicity of the two parameter model and the complexity of the experimental system.
	As such, the results of this fitting, shown in \reffig{fig:bcoef}, are intended only as a qualitative description.
	The form of $\beta'(\lambda)$ shows a decrease from the rapid decay experienced near resonance, to a minimum at $\sim$\SI{-1}{\tera\hertz} detuned, before smoothly increasing as the guide is detuned further.
	Is it possible to convert from the number loss rate rate $\beta'$ to the density loss rate $\beta$ by estimating the total volume of the atomic sample that we calculate from the Monte Carlo to be \SI{6e-6}{\centi\meter\cubed}.
	Using this value for the volume we calculate $\beta(\lambda)$ to be as low as \SI{5e-11}{\centi\meter\cubed\per\second} and as high as \SI{2e-9}{\centi\meter\cubed\per\second}, values that are comparable to values in the literature for similarly sized dipole traps \cite{Kuppens2000,Fuhrmanek2012}.
	
	As the density-dependent loss term restricts the usable optical depth of the system, the ability to select a wavelength for which the coefficient is minimized is of key importance.
	
\section{Performance and Limitations}
	We have demonstrated a highly efficient  mechanism for transferring cold atoms  into a waveguide.
	Most applications to which this extreme media could be targeted will depend on the achievable optical depth and coherence time of the media. 
	
	The highest optical depth attained in our system is \num{600\pm10} under typical operating conditions.
	The current limit to this performance is the due to the geometric mismatch between the guide beam and the MOT, with $\sim$\SI{3}{\percent} of the initial cloud coupled into the fibre.
	This fraction could be dramatically increased by using densifying techniques such as a spatial dark spot MOT, or changing to a high aspect ratio cigar shaped MOT.
	
	We have also investigated the dependence and dynamics of the loading process on wavelength and power of the guide - in this process we identified a density dependent atom-loss term that is likely associated with  photo-association.
	This process can result in an unwanted reduction in the number of loaded atoms, but can be alleviated by tuning the guide wavelength to avoid a photo-association transition.
	
	The atomic coherence time will be limited by the faster of two effects: decoherence induced by interaction between trapped atoms and the optical guide field, or the loss of atoms.
	Photon scattering in the current experiment limits the atomic coherence time to around \SI{200}{\micro\second}, although this rate could be much reduced by replacing the attractive guide with a hollow repulsive guide mode \cite{Poulin2011}.
	This trap geometry would confine atoms predominantly within the low intensity region, vastly reducing the scattering rate.
	We have simulated this situation, using a hollow trap of the same peak intensity as our Gaussian trap, to predict a time between photon scattering events of \SI{10}{\milli\second} - a hundred fold improvement over the current lifetime.
	An alternative approach to the same end would be the use of a weak far off resonance red-detuned trap; this approach has been shown to work by using an initial ensemble with a lower temperature or by only trapping the coldest fraction of the atomic cloud  \cite{Okaba2014, Xin2018}.
	In these weaker red or blue detuned traps, the background gas collisions will set a coherent light-atom interaction limit at \SI{130}{\milli\second}; however, even this limit is likely extended   by careful baking of the fibre to reduce outgassing.

\section{Conclusion}
	We have demonstrated a HC-PCF cold-atom-loading technique that is capable of loading \num{3e6} atoms into the fibre - a state of the art result.
	Aided by a robust Monte Carlo simulation we have explored the loading efficiency dependence on the parameters of the optical guide beam enabling us to generate confined atom samples with optical depths as high as \num{600\pm10}.
	Our rapid probing technique has allowed us to follow the evolution of atoms in real-time and we can use this to observe the fundamental de-coherence processes that apply to the approach.
	Our simulation allows us to show that one could maintain these high optical depths while extending the coherence time out to beyond \SI{10}{\milli\second}.

\begin{acknowledgments}
	We would like to thank the South Australian government for supporting this research through the PRIF program.
	
	This research was funded by the Australian government through the Australian Research Council (DE12012028).
	
	BMS acknowledged support from an ARC Discovery Early Career Researcher Award (DE170100752).

	We would like to thank Marcin Witkowski for his contribution to the design and construction of the scalar magnetic field cancellation coils.

\end{acknowledgments}

\appendix

\section{1: Optical Depth Calculations}
	The optical depth for a single atom in a gaussian probe field is given by
	
	\begin{align}
		\mathcal{D}^\mathrm{opt}_\mathrm{atom}(\Delta')&=\frac{2 \sigma_{\mathrm{D}_2}}{\pi w(z)^2} e^{\frac{-2(x^2+y^2)}{w(z)^2}}\times\nonumber\\
		&\sum_{F'}\frac{S_{3,F'}}{1+4{\left(\frac{\Delta'-\Delta_{3,F'}}{\Gamma}\right)}^2}
		\label{eqn:AtomOD}
	\end{align}
	
	where $\Delta'$ is the optical detuning in the atomic reference frame, $\sigma_{\mathrm{D}_2}$ is the photon scattering cross-section for the interrogated transition, $w(z)$ is the local waist of the probe, $S_{3,F'}$ are the hyperfine strength factors, and $\Delta_{3,F'}$ are the optical detunings of hyperfine levels.
	The velocity of the atom is accounted for by moving to the lab reference frame using $\Delta'=\Delta + v_z \frac{\omega_0}{c}$.
	
	The optical depth for an ensemble of atoms can be estimated by summing the optical depth over each atom, or equivalently, one can integrate over the spatial atom number density $n(\rho,z)$:
	
	\begin{equation}
		\mathcal{D}^\mathrm{opt} = \int_{0}^{L} \int_{0}^{r_\mathrm{core}} n(\rho,z) \mathcal{D}^\mathrm{opt}_\mathrm{atom} 2\pi\rho d\rho dL
		\label{eqn:FibreOD}
	\end{equation}
	
	where $L$ and $r$ are the length and radius of the ensemble, respectively.
	
	One can simplify this calculation by making the assumption that ensemble of $N_\mathrm{atom}$ atoms is uniformly populated over $L$, and that the radial dependence of the atom density follows a gaussian distribution with standard deviation $x_0$:
	
	\begin{equation}
		n(\rho,z) =
		\begin{cases}
		0, & \left|z\right|>L/2,\\
		N_\mathrm{atom} \frac{1}{L} \sqrt{\frac{2}{\pi x_0^2}} e^{-\frac{\rho^2}{2 x_0^2}}, &\left|z\right|<L/2.
		\end{cases}
		\label{eqn:AtomDensity}
	\end{equation}
	
	The integration now reduces to
	
	\begin{equation}
		\mathcal{D}^\mathrm{opt} = \eta N_\mathrm{atom} \frac{2 \sigma_{\mathrm{D}_2} S_{3,4}}{\pi w_0^2}
		\label{eqn:FibreODSimple}
	\end{equation}
	
	for the peak optical depth on the $F = 3 \rightarrow F = 4$ hyperfine transition, where $\eta$ is a coefficient that describes the level of spatial overlap between a cloud of atoms with gaussian radial density and uniform longitudinal density, and a gaussian intensity profile:
	
	\begin{equation}
		\eta = \frac{(w/2)^2}{x_0^2+(w/2)^2}.
		\label{eqn:Eta}
	\end{equation}

\section{2: Atom Coupling Dependence}
	The atom capture efficiency strongly depends on the geometries of the initial MOT and guide geometries.
	A brief algebraic analysis is done to determine the scaling of efficiency with the power and size of the guide beam.
	
	We first approximate the guide beam to be collimated through the length of the atom cloud, and we assume that the waist, $w_G$, is sufficiently small at the initial atom cloud location, $z_0$, that the cloud can be approximated as having a uniform column number density given by $\mathcal{N}$.
	At $z_0$ the trap depth in the radial direction can be described by
	
	\begin{equation}
		U_\mathrm{dip}\left(\rho\right) = U_0\left(P,\lambda\right) e^{-\frac{2\rho^2}{{w_G}^2}}
		\label{eqn:Udip}
	\end{equation}
	
	where $\rho$ is the radial coordinate, $P$ and $\lambda$ are the optical power and wavelength of the guide, and 
	
	\begin{align}
		U_0\left(P,\lambda\right) &= U_\mathrm{dip}\left(0\right)\\
		&= \alpha\left(\lambda\right) \frac{2 P}{\pi {w_G}^2}
		\label{eqn:U0}
	\end{align}
	
	is the peak trap depth, with wavelength dependence $\alpha\left(\lambda\right)$.
	
	We consider atoms to be trapped if the trap depth at their location is greater in magnitude than the kinetic energy of the atom, and negative in sign, i.e.

	\begin{equation}
		U_\mathrm{dip}\left(\rho\right) + KE_\mathrm{atom} < 0.
		\label{eqn:TrapCondition}
	\end{equation}
	
	The radius at which this is true for a fixed ensemble temperature is
	
	\begin{equation}
		\rho_\mathrm{trap} = w_G \sqrt{\frac{1}{2} \ln\left[-\frac{ U_0\left(P,\lambda\right)}{KE_\mathrm{atom}}\right]}
		\label{eqn:TrapRadius}
	\end{equation}
	
	and the number of trapped atoms thus given by
	
	\begin{align}
		N_\mathrm{atom} &= \pi {\rho_\mathrm{trap}}^2 \mathcal{N}\\
		&= \frac{1}{2} \pi  {w_G}^2 \mathcal{N} \ln\left[-\frac{ U_0\left(P,\lambda\right)}{KE_\mathrm{atom}}\right]\\
		&= \frac{1}{2} \pi  {w_G}^2 \mathcal{N} \ln\left[-\frac{2 P \alpha\left(\lambda\right)}{\pi {w_G}^2 KE_\mathrm{atom}}\right].
		\label{eqn:TrappedAtomNumber}
	\end{align}
	
	From this we can find the dependence of trapped atoms on guide power and waist:		
	
	\begin{align}
		\frac{d N_\mathrm{atom}}{d P} &= \mathcal{N} \frac{\pi {w_G}^2}{2 P},\\
		\frac{d N_\mathrm{atom}}{d {w_G}} &= \mathcal{N} \pi {w_G} \left(\ln\left[-\frac{2 P \alpha\left(\lambda\right)}{\pi {w_G}^2 KE_\mathrm{atom}}\right]-1\right).
		\label{eqn:DN}
	\end{align}

\begin{figure*}[ht]
		\centering
		\includegraphics[]{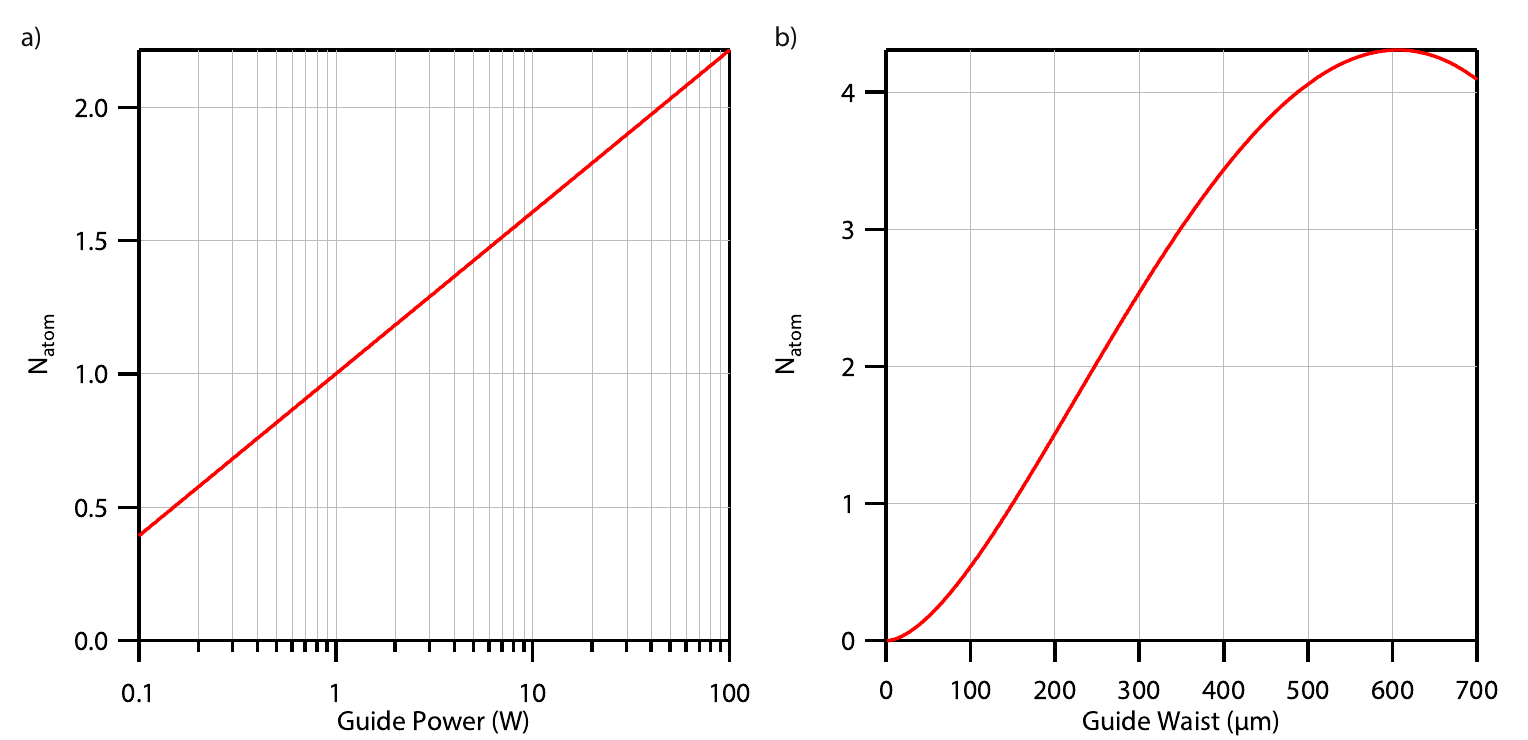}
		\caption{
			Calculated number of atoms trapped using Eqn. \ref{eqn:TrappedAtomNumber}, given relative to the atom number for $U_0\left(\SI{797.25}{\nano\meter},\SI{1}{\watt}\right)$ and $w_G=\SI{150}{\micro\meter}$, with
			a) $\lambda$ fixed at \SI{797.25}{\nano\meter}, $P$ varied, and
			b) $P$ fixed at \SI{1}{\watt}, with $w_G$ varied.
			\label{fig:Scaling}
		}
	\end{figure*}

	We conclude that the scaling of atom number with power offers strongly diminishing returns, while increasing the divergence of the guide or the distance from the fibre to the cloud could lift the trapped atom number as shown in \reffig{fig:Scaling}. 
		
\section{Monte Carlo Simulation}
	The local photon scattering rate for an alkali metal atom in a light field detuned from an optical transition by a distance larger than the hyperfine splitting, and less than the fine splitting, is given by
	
	\begin{equation}
		\Gamma_\mathrm{sc}(\mathbf{r})=\frac{\pi c^2 \Gamma}{2 \hbar \omega_0^3} \left(\frac{ 1}{\Delta_{\mathrm{D}_1}^2}+\frac{2}{\Delta_{\mathrm{D}_2}^2}\right) I(\mathbf{r})
		\label{eqn:GuideScattering}
	\end{equation}
	
	where $c$ is the speed of light, $\hbar$ is the reduced Planck constant, $w_0$ is the resonance angular frequency, $\Delta_{\mathrm{D}_1}$ and $\Delta_{\mathrm{D}_2}$ are the detunings from the centres of the D$_1$ and D$_2$ lines respectively, and $I(\mathbf{r})$ is the guide intensity.
	
	We integrate $\Gamma_\mathrm{sc}(\mathbf{r})$ while numerically solving the equations of motion for each atom to find the total number of scattering events up to a time t:
	
	\begin{equation}
		N_\mathrm{sc}(t) = \int_{0}^{t} \Gamma_\mathrm{sc}\left(\mathbf{r\left(\tau\right)}\right) d\tau.
		\label{eqn:ScatteringIntegral}
	\end{equation} 
	
	Each time $N_\mathrm{sc}{t}$ crosses an integer, we perturb the atom velocity with a kick, $\delta v_\mathrm{atom}$, of 
	
	\begin{equation}
		\delta v_\mathrm{atom}= \frac{\hbar \omega_\mathrm{G}}{\mathrm{m}_{\mathrm{Rb}} c} \hat{z} + \frac{\hbar \omega_{\mathrm{D}_1}}{\mathrm{m}_{\mathrm{Rb}} c} \hat{\zeta}
		\label{eqn:PhotonKick}
	\end{equation}
	
	where $\omega_\mathrm{G}$ is the optical angular frequency of the guide and $\hat{\zeta}$ is a unit vector randomly generated in spherical coordinates.
	The first term represents a velocity kick in the direction of the guide due to spontaneous absorption, and the second represents a velocity kick in a random direction due to spontaneous emission.
	
	The simulation also allows for background gas collisions by probabilistically eliminating atoms from the simulation.
	The probability that an atom does not experience a collision in the time frame $\delta t$ is given by
	
	\begin{equation}
		P_\mathrm{c} = e^{-\gamma_i \delta t }
	\end{equation}
	
	where $\gamma_i$ is atom loss rate either outside or inside of the fibre, $\gamma_\mathrm{chamber}$ of $\gamma_\mathrm{fibre}$.
	By sampling this probability at sufficiently fine time intervals during calculation the population decays by the appropriate rate.

%


\end{document}